\documentclass[doublecol]{epl2} 

\usepackage{ae} 
\usepackage[T1]{fontenc}
\usepackage[ansinew]{inputenc}
\usepackage{amsmath}
\usepackage{amssymb}
\usepackage{graphicx}
\usepackage[pdftex,usenames,x11names,svgnames,dvipsnames]{xcolor}
\usepackage[colorlinks]{hyperref}
\usepackage{cancel}
\usepackage{enumerate}

\usepackage[normalem]{ulem}
\usepackage{mathdots}
\usepackage{bm} 
\usepackage{mathrsfs} 
\usepackage{braket}

\usepackage{lscape} 
\newcommand{\ud}{\mathop{}\!\mathrm{d}} 

\title{On a no-go theorem for classical Maxwell-Lorentz electrodynamics in odd-dimensional worlds}
\shorttitle{On a non-go theorem for the classical electrodynamics in odd-dimensional worlds} 

\author{I. Aharonovich \inst{1} \and L. P. Horwitz\inst{1,2,3,4} }

\institute{                    
  \inst{1} Bar-Ilan University, Deppartment of Physics, Ramat Gan, Israel\\
  \inst{2} Tel-Aviv University, School of Physics, Ramat Aviv, Israel    \\
  \inst{3} College of Judea and Samaria, Ariel, Israel                   \\
  \inst{4} IYAR, Israel Institute for Advanced Research, Rehovot, Israel
}

\abstract{
A non-existence theorem of classical electrodynamics in odd-dimensional spacetimes is shown to be invalid.
The source of the error is pointed out, 
and is then demonstrated during the derivation of the fields generated by a uniformly moving point source.
}

\begin{document}

\maketitle

\section{Introduction}
In \cite{kosyakov_2011}, Kosyakov \emph{et al} have asserted a \emph{no-go} theorem for 
the existence of electromagnetic fields due to arbitrarily moving point sources
in odd-dimensional spacetimes. 
The proof showed that the potentials $A^{\mu}(x)$ due to such a source are \emph{pure gauge} $A^{\mu}(x) = \partial^{\mu} \chi(x)$ 
derived from some scalar function $\chi(x)$.
Therefore, any fields $F^{\mu\nu} = \partial^{\mu} A^{\nu} - \partial^{\nu} A^{\mu}$ derived from such potentials are \emph{identically zero}.

An error in the derivation, however, renders the conclusion false.
In the following, we show the exact point of error, and we demonstrate 
it along the a derivation of the fields generated by a uniformly moving point source.

\section{Maxwell electrodynamics in odd-dimensional spacetime}
Following \cite{kosyakov_2011}, we begin with the definition of the current $j^{\mu}(x)$ in 
some odd-N-dimensional spacetime. Given an arbitrary worldline $z^{\mu}(s)$ parameterized by 
proper time $s$, the \emph{covariant form} of the current is
\begin{align}
    \label{eq:def_current}
    j^{\mu}(x) 
    & = 
        e 
        \int_{-\infty}^{+\infty} 
            v^{\mu}(s) \delta^{N}(x - z(s)) 
            \,
            \ud s
\end{align}
where $v^{\mu}(s) = \dot{z}^{\mu} = \ud z^{\mu}/\ud s$.
In the generalized Lorentz gauge, the potentials $A^{\mu}(x)$ obey the $N$-dimensional wave equation:
\begin{align}
    \label{eq:wave_equation}
    \partial_{\nu} \partial^{\nu} A^{\mu}(x) & = j^{\mu}(x)
\end{align}

A solution for such an equation is given by     
\begin{align}
    \label{eq:A_mu_via_green_functions}
    A^{\mu}(x) 
    & = 
        \int
            G_{\text{ret}}(x - x') 
            j^{\mu}(x')
            \,
            \ud^{N}x'
\end{align}

The \emph{retarded} Green function $G_{\text{ret}}(x)$ is given by \cite{kosyakov_2011,galtsov_2002,kazinski_2002}
\begin{align}
    \label{eq:retarded_green_function}
    G_{\text{ret}}(x) 
    & = 
        \dfrac{(-1)^{n}}{\Omega_{N}}
        \theta(x_0)
        \Big( \dfrac{\ud}{\ud x^2} \Big)^{n}
            \dfrac{\theta(-x^2)} 
                  {\sqrt{-x^2}}
\end{align}
where $n$ is defined by $D = 2n+3$, and $x^2 = x_{\mu} x^{\mu}$.

Defining $R^{\mu}(s) = x^{\mu} - z^{\mu}(s)$, the potentials become
\begin{align}
    \label{eq:A_mu_2nd}
    A^{\mu}(x) 
    & = 
        (-1)^{n}
        \dfrac{e}{\Omega_{D}}
        \int_{-\infty}^{+\infty}
            v^{\mu}(s)
            \theta(R^{0})
            \Big( \dfrac{\ud}{\ud R^2} \Big)^{n}
            \dfrac{\theta(-R^2)}{\sqrt{-R^2}}
            \ud s
\end{align}

Then, the integration variable $s$ is substituted with $\lambda = -R^2$, 
and $\ud s/\ud \lambda = - 1/2 (R \cdot v)$, leading to 
\begin{align}
    \label{eq:A_mu_3rd}
    A^{\mu}(x) 
    & = 
        (-1)^{2n+1}
        \dfrac{e}{2 \Omega_{D}}
        \int_{0}^{\infty}
            \dfrac{ v^{\mu} }{R \cdot v}
            \Big( \dfrac{\ud}{\ud \lambda} \Big)^{n}
            \dfrac{\theta(\lambda)}{\sqrt{\lambda}}
            \ud \lambda
\end{align}

Now, the expression $v^{\mu}/(R \cdot v)$ is observed to be 
\mbox{$\partial^{\mu} \ln | R \cdot v|$}, and therefore
\begin{align}
    \label{eq:A_mu_4th}
    A^{\mu}(x) 
    & = 
        (-1)
        \dfrac{e}{2 \Omega_{D}}
        \int_{0}^{\infty}
            \dfrac{\partial}{\partial x_{\mu}} \ln | R \cdot v |
            \Big( \dfrac{\ud}{\ud \lambda} \Big)^{n}
            \dfrac{\theta(\lambda)}{\sqrt{\lambda}}
            \ud \lambda
\end{align}
Doing $n = D/2 - 3$ integration by parts and regularizing the boundary terms to be zero, 
we have
\begin{align}
    \label{eq:A_mu_5th}
    A^{\mu}(x) 
    & = 
        (-1)^{n+1}
        \dfrac{e}{2 \Omega_{D}}
        \int_{0}^{\infty}
            \dfrac{1}{\sqrt{\lambda}}
            \Big( \dfrac{\ud}{\ud \lambda} \Big)^{n}
            \dfrac{\partial}{\partial x_{\mu}} 
            \ln | R \cdot v |
            \ud \lambda
\end{align}
If one takes the partial derivative $\partial/\partial x_{\mu}$ outside the integral, one obtains
\begin{align}
    \label{eq:A_mu_6th}
    \begin{split}
        A^{\mu}(x) & = \partial^{\mu} \chi(x)
        \\
        \chi(x)    & = 
                (-1)^{n+1}
                \dfrac{e}{2 \Omega_{D}}
                \int_{0}^{\infty}
                    \dfrac{1}{\sqrt{\lambda}}
                    \Big( \dfrac{\ud}{\ud \lambda} \Big)^{n}
                    \ln | R \cdot v |
                    \ud \lambda
    \end{split}
\end{align}

The form of \eqref{eq:A_mu_6th} suggests that the potentials $A^{\mu}(x)$ are pure gauge.
However, \eqref{eq:A_mu_6th} is essentially \emph{incomplete}.
The very subtle key point lies in the substitution $\lambda = -R^2$, 
where the worldline 
function $z^{\mu}(s)$ and the associated velocity $v^{\mu}(s)$ are now functions of $\lambda$.
However, $\lambda = \lambda(x;s)$ is a function of $s$, the original integration variable, \emph{and the observation point $x^{\mu}$ as well}.
Now $s$ becomes a dependent variable, inverted from $\lambda(x;s)$ to be  $s = s(x;\lambda)$,
bringing \emph{explicit $x^{\mu}$ dependency} into $z^{\mu}, v^{\mu}$ and $R^{\mu}(x;\lambda) = x^{\mu} - z^{\mu}(x;\lambda)$.
Therefore, the corrected form of $v^{\mu}/R \cdot v$ is more complicated, i.e.,
\begin{align}
    \label{eq:fixed_log_derivative}
    \dfrac{v^{\mu}}{R \cdot v}
    & = 
        \dfrac{\partial}{\partial x_{\mu}} \ln|R \cdot v|
        -
        \dfrac{v^2 - R \cdot a}
              {R \cdot v}
        \dfrac{\partial s}{\partial x_{\mu}}
\end{align}
where $a^{\mu}(s) = \ud^2 z^{\mu}/\ud s^2$.
Equations \eqref{eq:A_mu_5th} and \eqref{eq:A_mu_6th} are therefore, \emph{incomplete}.
The second term in \eqref{eq:fixed_log_derivative} indicates that the additional terms 
are \emph{not in pure gauge form}, 
leading to the conclusion that $A^{\mu}(x)$ is \emph{not a pure gauge}.

This completes our argument.

\section{Fields due to a uniformly moving charge}
The worldline of a uniformly moving particle is given by
\begin{align}
    \label{eq:worldline_uniform_motion}
    z^{\mu}(s) & = b^{\mu} s 
\end{align}
where $v^{\mu} = b^{\mu}$ is \emph{constant velocity}, and $z^{\mu}(0)=0$.

Then, using \eqref{eq:A_mu_2nd}
\begin{align}
    \label{eq:A_uniform_motion}
    \begin{split}
        A^{\mu}(x) 
        & = 
            (-1)^{n}
            \dfrac{e}{\Omega_{D}}
            \int_{-\infty}^{+\infty}
                b^{\mu}
                \theta(t - b^{0}s)
                \cdot
        \\
        & \qquad \qquad 
                \cdot 
                \Big( \dfrac{\ud}{\ud a} \Big)^{n}
                \dfrac{\theta(- (x-bs)^2 + a)}{\sqrt{-(x-bs)^2+a}} 
                \ud s
                \Big|_{a=0}
    \end{split}
\end{align}
Now:
\begin{align}
    \label{eq:lambda_uniform_motion}
    \lambda & = - x^2 + 2 s (b \cdot x) - s^2b^2  = -x^2 + 2s(b\cdot x) + s^2
\end{align}
where we have taken the \emph{on-shell} condition $b^2 = -1$.
The support in the past light-cone leads to the upper bound on $s$
\begin{align}
    \label{eq:A_uniform_motion_2}
    \begin{split}
        A^{\mu}(x) 
        & = 
            (-1)^{n}
            \dfrac{e b^{\mu}}{\Omega_{D}}
            \Big( \dfrac{\ud}{\ud a} \Big)^{n}
            \int_{-\infty}^{s_{-}}
                \dfrac{1}{\sqrt{-(x-bs)^2+a}} 
                \ud s
                \Big|_{a=0}
    \end{split}
\end{align}
where
\begin{align}
    s_{-} & = - b \cdot x - \sqrt{(b \cdot x)^2 + x^2} 
\end{align}

Defining:
\begin{align}
    r^2 & = (b \cdot x)^2 + x^2
\end{align}
and changing variables from $s$ to $\theta$:
\begin{align}
    s + b\cdot x & = - r \cosh(\theta),
\end{align}
we then have
\begin{align}
    \label{eq:A_uniform_motion_3}
    \begin{split}
        A^{\mu}(x) 
        & = 
            (-1)^{n+1}
            \dfrac{e b^{\mu}}{\Omega_{D}}
            \cdot 
        \\
        & \qquad \qquad 
            \cdot
            \Big( \dfrac{\ud}{\ud a} \Big)^{n}
            \int_{-\infty}^{0}
                \dfrac{r \sinh(\theta) \ud \theta}{\sqrt{r^2 \cosh^2(\theta)- r^2 + a }} 
                \Big|_{a=0}
        \\
        & = 
            (-1)^{n+1}
            (-1)^{n} 
            \dfrac{(2n-1)!!}{2^{n} r^{2n}} 
            \dfrac{e b^{\mu}}{\Omega_{D}}
            \int_{-\infty}^{0}
                \dfrac{\ud \theta}{ \sinh^{2n}(\theta) } 
    \end{split}
\end{align}
The integral $\int \ud \theta/\sinh^{2n}(\theta) $ is now a constant (the diverging bound at $0$ is regularized away)
and therefore:
\begin{align}
    \label{eq:A_mu_uniform_motion_generic}
    A^{\mu}(x) & = C_{n,D} \dfrac{e b^{\mu}}{((b \cdot x)^2 + x^2)^{n}}
\end{align}
where $C_{n,D}$ depends only on dimensionality.
The fields derived from \eqref{eq:A_mu_uniform_motion_generic} 
are
\begin{align}
    F^{\mu \nu}(x)
    & = 
        2 e
        C_{n,D} 
        \dfrac{b^{\mu} x^{\nu}  - b^{\nu} x^{\mu} }
              {((b \cdot x)^2 + x^2)^{n+1}}
\end{align}
and are therefore non-zero.

One can see from \eqref{eq:lambda_uniform_motion}
that
\begin{align}
    s(x;\lambda) 
    & = 
        - (b \cdot x) - \sqrt{(b \cdot x)^2 + x^2 - \lambda}
\end{align}
and therefore, $\partial s/\partial x^{\mu} \neq 0$. 
Thus, if one changes variables from $s$ to $\lambda(x;s) = -R^2(s)$, then
\begin{align}
    \label{eq:A_uniform_motion_lambda_substitution_1}
    \begin{split}
        A^{\mu}(x) 
        & = 
            \dfrac{(2n-1)!!}{2^{n}}
            \dfrac{e b^{\mu}}{\Omega_{D}}
            \int_{-\infty}^{s_{-}}
                \dfrac{1}{\lambda^{n+1/2}}
                \dfrac{\ud \lambda}{R(\lambda) \cdot b}
    \end{split}
\end{align}
Now, $R_{\mu}(\lambda) = x_{\mu} - b_{\mu} s(x;\lambda)$, and clearly
\begin{align}
    \dfrac{b^{\mu}}{R(\lambda) \cdot b}
    & =
        \dfrac{\partial}{\partial x_{\mu}} \ln |R \cdot b|
        +
        \dfrac{b^2}{R \cdot b} \dfrac{\partial s(x;\lambda)}{\partial x_{\mu}}
\end{align}

\section{Conclusion}
The no-go theorem for the non-existence of electromagnetic fields due to a point source in odd-dimensional spacetimes is proved to be incorrect.
The exact source of error was due to overlooking the somewhat hidden explicit $x^{\mu}$ dependence of the worldline functions $z^{\mu}(s) \to z^{\mu}(x;\lambda)$,
which breaks the pure gauge construct.
This was explicitly demonstrated in the uniform motion case, where the inversion $s(x;\lambda)$ is relatively easy to achieve.

Finally, we note there already exists a substantial body of theoretical work that supports electrodynamics in odd-dimensional spacetimes, 
cf. \cite{land_2011,galtsov_2002,kazinski_2002,jigal_2006} and references therein.

\acknowledgments
We wish to thank Prof. B. P. Kosyakov for the fruitful discussions and for showing the true spirit of science.

\end{document}